\documentclass[oupdraft]{bio}
\usepackage{url}

\usepackage{algorithm,algorithmicx}
\usepackage{algpseudocode}
\algdef{SE}[SUBALG]{Indent}{EndIndent}{}{\algorithmicend\ }%
\algtext*{Indent}
\algtext*{EndIndent}

\history{Received September 22, 2021;
revised October XX, 2021;
accepted for publication November XX, 2021}

\begin{document}

\title{Variable Selection in GLM and Cox Models with Second-Generation P-Values}

\author{YI ZUO$^\ast$, THOMAS G. STEWART, JEFFREY D. BLUME\\[4pt]
\textit{Department of Biostatistics, Vanderbilt University, Nashville TN, USA}
\\[2pt]
{yi.zuo@vanderbilt.edu}}

\markboth%
{Yi Zuo and others}
{Variable Selection in GLM and Cox Models with SGPV}

\maketitle

\footnotetext{To whom correspondence should be addressed.}

\begin{abstract}
{Variable selection has become a pivotal choice in data analyses that impacts subsequent inference and prediction. In linear models, variable selection using Second-Generation P-Values (SGPV) has been shown to be as good as any other algorithm available to researchers. Here we extend the idea of Penalized Regression with Second-Generation P-Values (ProSGPV) to the generalized linear model (GLM) and Cox regression settings. The proposed ProSGPV extension is largely free of tuning parameters, adaptable to various regularization schemes and null bound specifications, and is computationally fast. Like in the linear case, it excels in support recovery and parameter estimation while maintaining strong prediction performance. The algorithm also preforms as well as its competitors in the high dimensional setting ($n>p$). Slight modifications of the algorithm improve its performance when data are highly correlated or when signals are dense. This work significantly strengthens the case for the ProSGPV approach to variable selection. }
{variable selection; second-generation p-values; generalized linear model; Cox model}
\end{abstract}

\section{Introduction}
\label{sec1}

Generalized linear models (GLM)  (\citealp{nelder1972generalized}) and the Cox proportional hazards model (\citealp{cox1972regression}) are by now essential tools for regression modeling beyond linear models. GLMs can accommodate binary outcomes (disease status), count data (number of readmissions), and other type of continuous outcomes, while Cox models allow for time-to-event data (survival). Both are widely used in daily practice. While the magnitude of data – and their modeling options – has grown exponentially over the past decade, typically only a small finite set of features remains of interest to the researcher (\citealp{wainwright2009sharp,loh2017support,gao2020fundamental}). As such, variable selection remains a crucial task because it often leads to better risk assessment, parameter estimation and model interpretation (\citealp{shmueli2010explain}). Sometimes, the number of variables, say $p$, can be (much) larger than the number of observations $n$ (e.g., see microarray, proteomic, and single nucleotide polymorphisms (SNPs) data). In those cases, penalized likelihood methods are often employed to encourage sparsity so that subsequent statistical inference is reliable.

In this paper, we extend the penalized regression with second-generation p-values (ProSGPV) (\citealp{zuo2021variable}) from the linear regression setting to GLMs and Cox models. The key ideas remain the same, although their implementation requires special attention outside of the linear case. First, the algorithm identifies a candidate set that is likely to contain the true signals. Then is proceeds to screen out noise variables using second-generation p-value (SGPV) methodology (\citealp{blume2018second,blume2019introduction}). Traditional \textit{p}-values only measure the agreement between data and a point null hypothesis. By contrast, SGPVs measure the strength of an effect by comparing its entire uncertainty interval to a pre-specified interval null hypothesis. This idea of incorporating uncertainty into the variable selection procedure is novel and readily generalizable to different settings. We show below by simulation that the ProSGPV algorithm has very competitive support recovery, parameter estimation, and prediction performance in comparison to the current standards across Logistic, Poisson, and Cox models. Moreover, ProSGPV is largely insensitive to hard-to-interpret tuning parameters. The algorithm’s framework has a flexible setup and is robust to modifications to many steps. 

The rest of the paper is organized as follows. Section \ref{sec2} describes the current landscape of pertinent variable selection methods for models of non-linear outcomes. Section \ref{sec3} details the generalized ProSGPV variable selection algorithm. Section \ref{sec4} presents simulation studies comparing inference and prediction performance of the aforementioned methods when n>p and in high dimensional settings when $p>n$. Section \ref{sec5} illustrates the ProSGPV algorithm using a real-world example. Section \ref{sec6} summarizes key findings and discusses limitations.

\section{Current landscape}
\label{sec2}

Let $\{(\boldsymbol X_i,Y_i)\}^n_{i=1}$ be a collection of $n$ independent observations, where $Y_i$ are independently observed response values given the $p$-dimensional predictor vector $\boldsymbol X_i$. With a canonical link, the conditional distribution of $Y_i$ given $\boldsymbol X_i$ belongs to an exponential family with density function $f(Y_i\vert \theta_i)\propto \exp(Y_i\theta_i-b(\theta_i))$, where $\theta_i=\boldsymbol Z^T_i\boldsymbol \beta$ with $\boldsymbol Z_i=(1,\boldsymbol X^T_i)^T$, $\boldsymbol\beta=(\beta_0,\beta_1,...,\beta_p)^T$, and $b(\theta)$ is assumed to be twice continuously differentiable with positive $b^{''}(\theta)$. The loss function (negative log-likelihood) of a GLM model is 
\begin{equation}
L(\boldsymbol\beta)=-\frac{1}{n}\sum_{i=1}^n[Y_i\boldsymbol Z_i^T\boldsymbol\beta-b(\boldsymbol Z_i^T\boldsymbol \beta)]
\label{glm-loss}
\end{equation}
Penalized likelihood methods optimize a loss function that is a slight variation on Equation (\ref{glm-loss}), which is 
\begin{equation}
L_\lambda(\boldsymbol\beta)=-\frac{1}{n}\sum_{i=1}^n[Y_i\boldsymbol Z_i^T\boldsymbol\beta-b(\boldsymbol Z_i^T\boldsymbol \beta)]+p_\lambda(\boldsymbol\beta)
\label{glm-p-loss}
\end{equation}
where $p_\lambda(\boldsymbol\beta)$ is the penalty function that involves a tuning parameter $\lambda$ and the coefficients $\boldsymbol\beta$. In the case of a Cox model, the log partial likelihood is used for the likelihood and penalized by $p_\lambda(\boldsymbol\beta)$. Almost all variable selection methods in this setting amount to selecting a particular penalty function.

\subsection{Lasso}

Lasso (\citealp{tibshirani1996regression,tibshirani1997lasso}) is one of the most widely used variable selection methods. It introduces an $\ell_1$ penalty to the loss function and shrinks effects towards zero. The sparsity induced by the $\ell_1$ penalty facilitates variable selection. The lasso GLM can be written as 
\begin{equation}
L^{\text{lasso}}_\lambda(\boldsymbol\beta)=-\frac{1}{n}\sum_{i=1}^n[Y_i\boldsymbol Z_i^T\boldsymbol\beta-b(\boldsymbol Z_i^T\boldsymbol \beta)]+\lambda\Vert \boldsymbol\beta\Vert_1
\end{equation}
where $\Vert \cdot\Vert_1$ is the $\ell_1$-norm.

While parameter and risk estimation properties of lasso has been studied in high dimensional GLM under Lipschitz loss function(\citealp{meier2008group}) as well as usual quadratic loss in GLM (\citealp{wang2015convergence}), it is hard to find the ideal $\lambda$ that achieves excellent support recovery with excellent parameter estimation properties in practice.

\subsection{BeSS}

Best subset selection (BSS) evaluates all sub-models in the feature space and searches for the one with the smallest loss criterion. Mathematically, BSS uses an $\ell_0$ penalty in the loss function. The BSS loss function can be written as 
\begin{equation}
L^{\text{BeSS}}_\lambda(\boldsymbol\beta)=-\frac{1}{n}\sum_{i=1}^n[Y_i\boldsymbol Z_i^T\boldsymbol\beta-b(\boldsymbol Z_i^T\boldsymbol \beta)]+\lambda\Vert \boldsymbol\beta\Vert_0
\end{equation}
where $\Vert \cdot\Vert_0$ is the number of non-zero element in a vector. The BSS problem is nonconvex and NP-hard (\citealp{natarajan1995sparse}). A fast implementation of BSS for GLM/Cox models, called BeSS, was recently proposed \citealp{wen2020bess}). When the golden section primal-dual active set (GPDAS) algorithm is used, there is no need to pre-specify the size of the model, and BeSS is still able to identify the best sub-model very quickly even when the number of features p is around 10,000. However, little is known about the asymptotic inference and prediction properties of BeSS solutions using GPDAS algorithm.

\subsection{ISIS}

Sure independence screening (SIS) was first proposed in linear regression (\citealp{fan2008sure}), and later extended to GLM models (\citealp{fan2010sure}). SIS works by ranking the maximum marginal likelihood estimators in GLM/Cox - the univariate estimates - and then selecting the top variables for the final model. However, SIS can miss variables that are marginally and weakly correlated with the response, but highly important in a joint sense; SIS’s dependence on marginal models may also cause it to rank some jointly unimportant variables too high. These and related issues were partially addressed by modifying the SIS algorithm, such as in multi-stage SIS, which is SIS followed by second-stage variable selection using lasso or smoothly clipped absolute deviation (SCAD), and iterative SIS (ISIS). The surprisingly simple ISIS often leads to good support recovery and parameter estimation. However, in order for ISIS to be variable selection consistent, the true signals should be sparse and not small (\citealp{fan2010sure}).

\subsection{SGPV}

Second-generation p-values (SGPV), denoted as $p_\delta$, were proposed for use in high dimensional multiple testing contexts (\citealp{blume2018second,blume2019introduction}). SGPVs attempt to resolve some of the deficiencies of traditional \textit{p}-values by replacing the point null with a pre-specified interval null $H_0=[-\delta,\delta]$. The interval null $H_0$ is the set of effects that are scientifically indistinguishable or immeasurable from the point null hypothesis, due to limited precision or practicality. The ‘effects’ are typically measured as log odds ratios (logistic regression), log rate ratios (Poisson regression), and log hazards ratios (Cox regression). SGPVs are loosely defined as the fraction of data-supported hypotheses that are also null, or nearly null, hypotheses. Formally, let $\theta$ be a parameter of interest, and let $I=[\theta_l, \theta_u]$ be the interval estimate of $\theta$ whose length is given by $|I|=\theta_u-\theta_l$. If we denote the length of the interval null by $|H_0|$, then the SGPV $p_\delta$ is defined as
\begin{equation}
p_\delta=\frac{|I\cap H_0|}{|I|}\times \max\left\{\frac{|I|}{2|H_0|},1\right\}  
\end{equation}
where $I\cap H_0$ is the intersection of two intervals. The correction term $\max\{|I|/(2|H_0|),1\}$ fixes the problem when the interval estimate is very wide, i.e., when $|I|>2|H_0|$. In that case, the data are effectively inconclusive and the main quantity $|I\cap H_0|/|I|=|H_0|/|I|$ does not properly reflect this inconclusive nature of the data. As such, SGPV indicate when data are compatible with null hypotheses ($p_\delta = 1$), or with alternative hypotheses ($p_\delta = 0$), or when data are inconclusive ($0 < p_\delta < 1$). 

By design, SGPV emphasize effects that are clinically meaningful by exceeding a pre-specified null level. Empirical studies have shown that SGPV can identify feature importance in linear models (\citealp{blume2018second,blume2019introduction,zuo2021variable}), and we will show below that SGPV can also be used for variable selection in GLM/Cox models. 

\section{ProSGPV algorithm in GLM/Cox}
\label{sec3}

The penalized regression with second-generation p-values (ProSGPV) algorithm was proposed for linear regression (\citealp{zuo2021variable}), but it is readily adapted to accommodate Logistic, Poisson, and Cox regressions.

\subsection{Steps}

The steps of the ProSGPV algorithm are shown below in Algorithm \ref{Algorithm1}.

\begin{algorithm}[H]
\caption{\label{Algorithm1}The ProSGPV algorithm}
\begin{algorithmic}[1]
    \Procedure{ProSGPV}{$\boldsymbol X$, $\boldsymbol Y$}
    \State \textbf{Stage one}: Find a candidate set 
    \Indent
    \State Fit a lasso and evaluate it at $\lambda_{\text{gic}}$ 
    \State Fit GLM/Cox models on the lasso active set  
    \EndIndent
    \State \textbf{Stage two}: SGPV screening 
    \Indent
    \State Extract the confidence intervals of all variables from the previous step
    \State Calculate the mean coefficient standard error $\overline{SE}$
    \State Calculate the SGPV for each variable where $I_j=\hat\beta_j\pm1.96\times SE_j$ and $H_0=[-\overline{SE},\overline{SE}]$
    \State Keep variables with SGPV of zero
    \State Refit the GLM/Cox with selected variables 
    \EndIndent
    \EndProcedure
\end{algorithmic}
\end{algorithm}

In the first stage, lasso is evaluated at the $\lambda$ that minimizes the generalized information criterion (\citealp{fan2013tuning}). The second step is a fully relaxed lasso (\citealp{meinshausen2007relaxed}) on variables with non-zero coefficients in the first step. The second stage uses SGPVa to screen variables. Throughout this paper, SGPV are calculated based on 95\% confidence intervals, though other interval choices would also work. The ProSGPV algorithm has been implemented in the ProSGPV R package, which is available from the Comprehensive R Archive Network (CRAN) at \url{https://CRAN.R-project.org/package=ProSGPV}.

\subsection{Flexibility of the algorithm}

\begin{figure}[!p]
\centering\includegraphics[width=\textwidth]{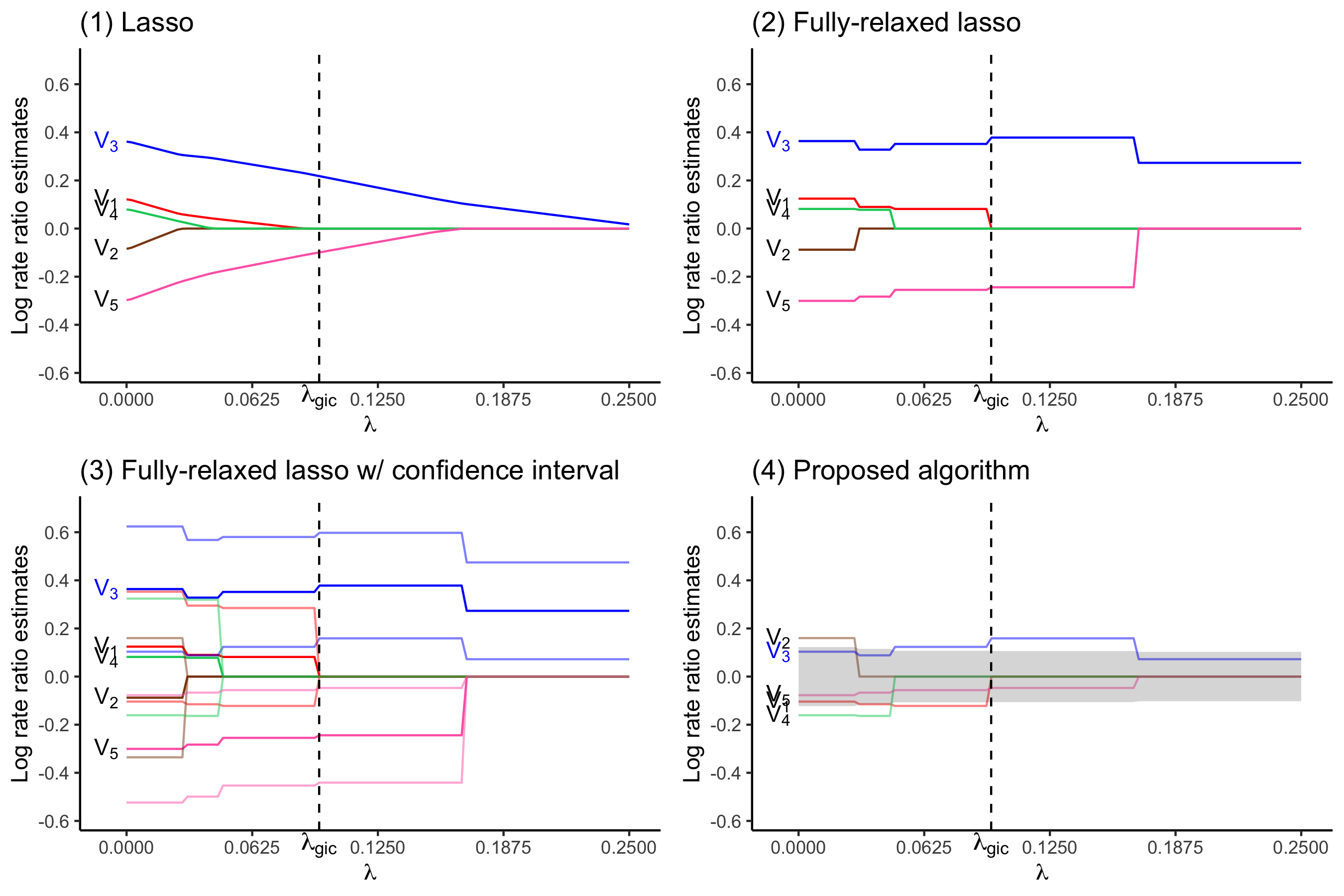}
\caption{How ProSGPV works in a Poisson regression. The true data generating model contains only $V_3$. (1) presents the lasso solution path. (2) shows the fully relaxed lasso path. (3) shows the fully relaxed lasso paths with their 95\% confidence intervals (in lighter color). (4) illustrates the ProSGPV selection path. The shaded area is the null region; the colored lines are each 95\% confidence bound that is closer to the null region. }
\label{Fig1}
\end{figure}

In ProSGPV, lasso is used in the first-stage to leverage its quick computation speed. Other dimension reduction methods, e.g. elastic net or adaptive lasso or even SIS, once properly tuned, should also lead to good inference and prediction performance in the ProSGPV algorithm as well (this tangent idea will be evaluated in a separate venue). Moreover, ProSGPV only requires that  the choice of $\lambda$ for the first-stage lasso be only in a certain range. This is because the feature space for the fully relaxed lasso is discrete and so there exists a range of $\lambda$s that lead to the same fully relaxed lasso model (See Figure \ref{Fig1} for details). In practice we suggest using the $\lambda$ that minimizes the generalized information criterion. Supplementary Figure 1 shows that as long as $\lambda$ falls into a reasonable range, the support recovery of ProSGPV is consistently good. Although using exactly $\lambda_\text{gic}$ may be preferable in high dimensional settings where $p>n$ (\citealp{fan2013tuning}). 

Another flexible aspect of the ProSGPV algorithm is the choice of the null bound in the second stage. This can be thought of as a tuning parameter, but one that is less sensitive. For a default, we choose the average standard error (SE) of coefficient estimates. For example, in Logistic regression, the null bound would be set at the SE of log odds ratio. Other forms of the null bound may lead to decent support recovery as well, although the Type I/Type II error tradeoff needs to be formally assessed. In Supplementary Figure 2, we compared the performance of ProSGPV using various null bounds such as $\overline{SE}$, $\overline{SE}*\sqrt{\log(n/p)}$, $\overline{SE}/\sqrt{\log(n/p)}$, $\overline{SE}*\sqrt{n/p}/2$, and the constant zero, in logistic regression. $\overline{SE}*\sqrt{n/p}/2$ is roughly constant when $n$ increases with fixed $p$ and is the largest among all bounds. When the null bound is multiplied by $\sqrt{\log(n/p)}$,  or divided by $\sqrt{\log(n/p)}$, the bound is inflated or deflated at each $n$. When null bound is zero, it’s effectively selecting variables using the traditional \textit{p}-values. The original null bound achieves variable selection consistency at the fastest rate. The other null bounds either are not variable selection consistent when $n$ is large (0 and $\overline{SE}/\sqrt{\log(n/p)}$), or achieve the consistency at a slower rate ($\overline{SE}*\sqrt{\log(n/p)}$ and $\overline{SE}*\sqrt{n/p}/2$). Hence our recommendation for using the original null bound.

An important note is that logistic regression fitted by maximum likelihood can suffer from inflated parameter estimation bias due to complete/quasi-complete separation when n is small (\citealp{albert1984existence, rahman2017performance}). A Jeffreys-prior penalty (\citealp{kosmidis2021jeffreys}) was included in ProSGPV to address this issue. A comparison of the two models is shown in the Supplementary Figure 3. When signals are dense and $n$ is small, Logistic regression with Jeffreys prior has lower parameter estimation bias and better prediction accuracy than Logistic regression fitted by maximum likelihood.

\subsection{Solution}

The solution to the ProSGPV algorithm $\hat{\boldsymbol\beta}^{pro}$ is 
\begin{equation}
\begin{gathered}
\hat{\boldsymbol\beta}^{\text{pro}}=\hat{\boldsymbol\beta}^{\text{gc}}_{|S}\in\mathbb R^p, \text{ where } \\
S=\{k\in C: |\hat\beta_k^{\text{gc}}|>\lambda_k \}, C=\{j\in\{1,2,...,p\}: |\hat{\beta}^{\text{lasso}}_j|>0\}
\end{gathered}
\end{equation}
where $\hat{\boldsymbol\beta}^\text{gc}_{|S}$ is a vector of length $p$ with non-zero elements being the GLM/Cox coefficient estimates from the model with variables only in the set $S$. $S$ is the final selection set and $C$ is the candidate set from the first-stage screening. The coefficients are log odds ratios in Logistic regression, log rate ratios in Poisson regression, and log hazard ratios in Cox regression. Note that the cutoff $\lambda_j=1.96*SE_j+\overline{SE}$.

Like the linear setting, ProSGPV can be seen as a hard thresholding function. In the saturated GLM/Cox models one can often order coefficients such that true signals all have greater coefficient estimates than noise variables do. The cutoff $\lambda_j$ aims to separate signal variables from noise variables. The only routine exception to this ordering is when the data have weak signals or high correlation, and in this case, no algorithms can fully recover the true support. (\citealp{zhao2006model,wainwright2009information})

\subsection{Example}

An example of how ProSGPV works in Poisson regression is shown below. There are five explanatory variables $(\boldsymbol V_1,..,\boldsymbol V_5 )\in \mathbb R^{100\times 5}$ and the covariance matrix is autoregressive with autocorrelation being 0.5 and standard deviation of 1. The response $\boldsymbol Y\in \mathbb R^{100\times 1}$ is simulated from Poisson distribution with mean of $\exp(\beta_3\boldsymbol V_3)$  where $\beta_3=0.25$. The goal is support recovery, i.e. to discover which variable is truly associated with the outcome. Figure \ref{Fig1} shows how ProSGPV succeeds while lasso and fully relaxed lasso fail at $\lambda=\lambda_\text{gic}$ by selecting $\boldsymbol V_3$ and $\boldsymbol V_5$.

\section{Simulation studies}
\label{sec4}

Extensive simulation studies were conducted to evaluate inference and prediction performance of the ProSGPV algorithm compared to existing standard methods, in both traditional $n>p$ and high-dimensional $p>n$ settings. The models under consideration include Logistic regression, Poisson regression, and Cox proportional hazards regression.  

\subsection{Design}

Given sample size $n$, dimension of explanatory variables $p$, sparsity level $s$, true coefficient vector $\boldsymbol \beta_0\in\mathbb R^p$, autocorrelation level $\rho$ and standard deviation $\sigma$ in the covariance matrix, the simulation steps are as follows.  

Step 1: Draw $n$ rows of the input matrix $\boldsymbol X\in\mathbb R^{n\times p}$ i.i.d. from $N_p (\boldsymbol 0,\Sigma)$, where $\Sigma \in \mathbb R^{p\times p}$ has entry $(i,j)$ equal to $\sigma^2 \rho^{|i-j|}$  

Step 2: For Logistic regression, generate the linear vector $\boldsymbol z=\boldsymbol{X\beta}_0\in\mathbb R^{n\times 1}$, where the intercept is 0, and generate response $\boldsymbol Y\in\mathbb R^{n\times 1}$ from Bernoulli distribution with probability $\boldsymbol{pr}=1/(1+\exp(-\boldsymbol z))$ For Poisson regression, generate $\boldsymbol z\in\mathbb R^{n\times 1}$ as before, and generate response $\boldsymbol Y\in\mathbb R^{n\times 1}$ from Poisson distribution with mean $\exp(\boldsymbol z)$. For Cox regression, generate time to event from Weibull distribution with scale $\lambda_\text{weibull}=2$, shape $k=1$, and mean $\exp(\boldsymbol z)$. generate censoring time from Exponential distribution with rate $\tau=0.2$

Step 3: Run ProSGPV, lasso, BeSS, and ISIS, and record whether each algorithm recovers exactly, and only, the true underlying features; compute the proportion of correctly captured true features (power) and the proportion of incorrectly captured noise features (Type I Error rate); compute the false discovery proportion (pFDR) and false non-discovery rate (pFNDR); compute the mean absolute bias in parameter estimation and prediction area under the curve (AUC) in a separate test set except for Cox model

Step 4: Repeat the previous steps 1000 times and aggregate the results over iterations 

$\boldsymbol\beta_0$ has $s$ non-zero values equally spaced between $\beta_l$ and $\beta_u$, at random positions, and the remaining coefficients are set to zero. The non-zero coefficients are half positive and half negative. $\rho$ is 0.35 and $\sigma$ is 2. Additional simulation parameters are summarized in Table \ref{Table1}.

The ProSGPV algorithm was implemented using the ProSGPV package; lasso was implemented using glmnet package and was evaluated at $\lambda_\text{min}$; BeSS was implemented using BeSS package;  and ISIS was implemented using SIS package. The code to replicate all results in this paper is available at \url{https://github.com/zuoyi93/r-code-prosgpv-glm-cox}.

\subsection{Results and findings}

\begin{figure}[!p]
\centering\includegraphics[width=\textwidth]{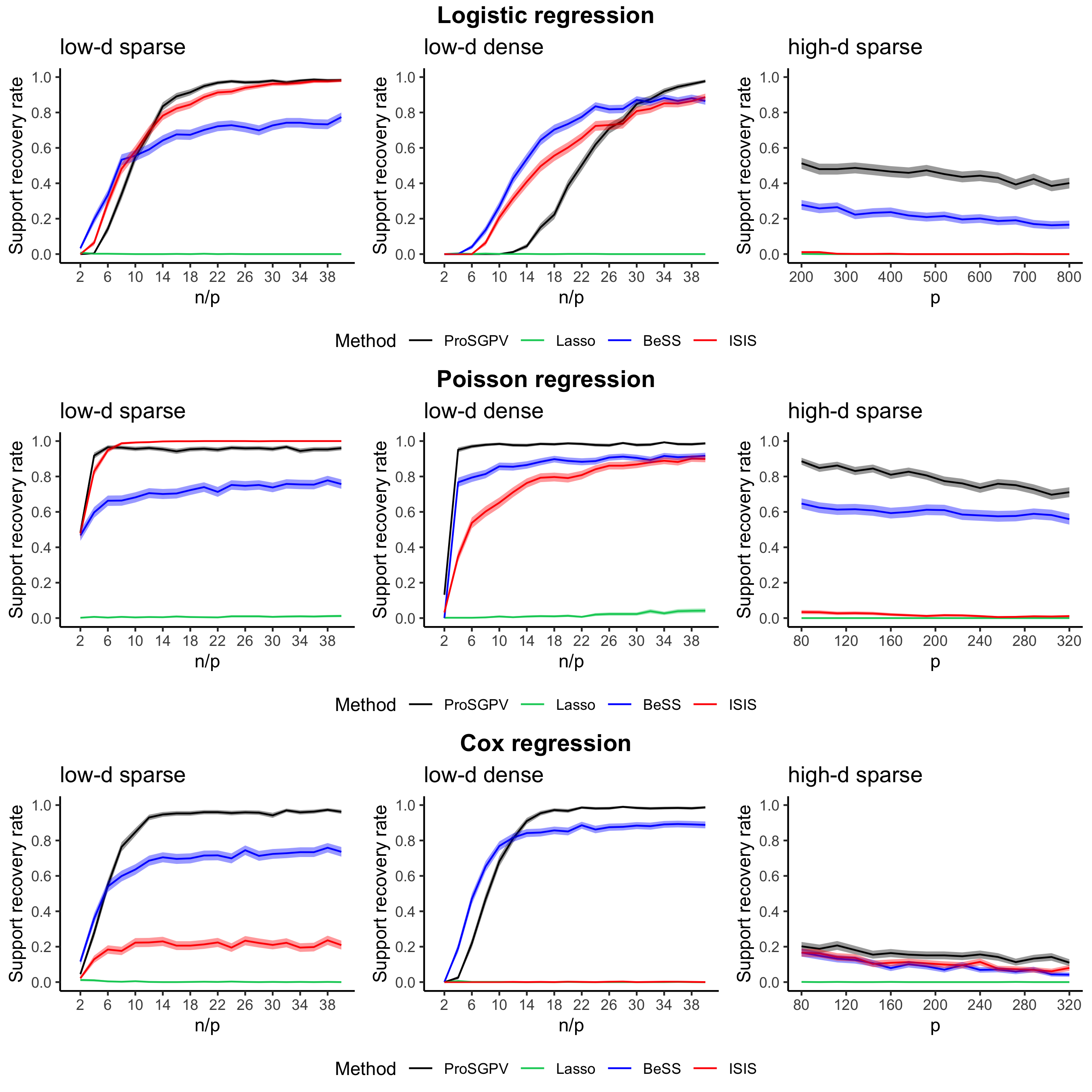}
\caption{Comparison of capture rate of the exact true model: mean rates surrounded by 95\% Wald confidence intervals over 1000 simulations }
\label{Fig2}
\end{figure}

\begin{figure}[!p]
\centering\includegraphics[width=\textwidth]{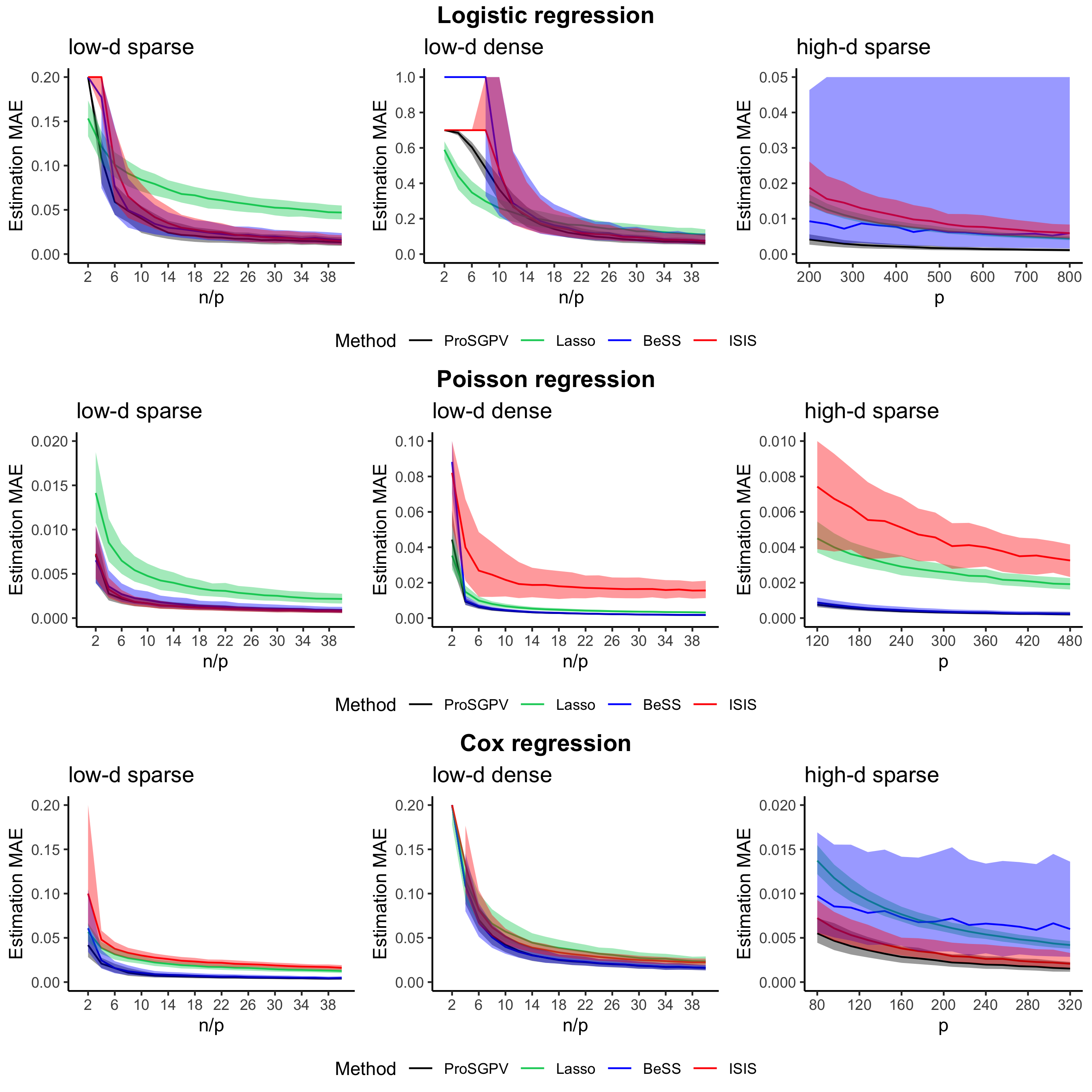}
\caption{Comparison of parameter estimation: medians of mean absolute error in parameter estimation surrounded by first and third quartiles over 1000 simulations}
\label{Fig3}
\end{figure}

The capture rates of the exact true model are shown in Figure \ref{Fig2}. The average coefficient estimation error is compared in Figure \ref{Fig3}. The prediction accuracy is displayed in Figure \ref{Fig4}. Power (the fraction of true features that were captured) and Type I Error rate are compared in Supplementary Figure 4. False discovery proportion (pFDR) and false non-discovery proportion (pFNR) are compared in Supplementary Figure 5. Computation time is shown in Supplementary Figure 7.

In Figure \ref{Fig2}, capture rates of the exact true model are compared under Logistic regression, Poisson regression, and Cox regression with various $(n,p,s)$ combinations. Overall, ProSGPV has the highest capture rate in almost all scenarios, particularly in high-dimensional setting where $p>n$. But there are exceptions. For example, ProSGPV has a low capture rate in logistic regression when signals are dense and $n/p$ is small (low-d dense column in Figure \ref{Fig2}). In that setting, BeSS has the best support recovery among the four, with ISIS not far behind. Interestingly, ISIS recovers support well in certain low-dimensional models but fails in high dimensional settings. Here, Lasso is evaluated at $\lambda_\text{min}$ which is optimized for prediction and completely fails in the support recovery task.

We further investigated the support recovery results by comparing their Type I error rates and power in Supplementary Figure 4 and false discovery proportion (pFDR) and false non-discovery proportion (pFNR) in Supplementary Figure 5. In general, ProSGPV selects most signal variables and seldom includes noise variables. However, ProSGPV has a low power in a dense Logistic regression when $n$ is small. BeSS selects more signals than ProSGPV at the cost of including more noise variables. ISIS selects slightly fewer signal variables and more noise variables, which becomes problematic for high-dimensional support recovery. Interestingly, lasso fails the support recovery task because it selects a large model that, while including almost all signal variables, has numerous noise variables. We see that ProSGPV has superior or comparable support recovery performance as the other standard procedures with the exception of the low-dimensional dense signal logistic regression. This happens in part because, in that case, the null bound is slightly higher than the noise level and ProSGPV includes too many noise variables. This can be fixed by replacing the constant null bound with an adjusted null bound based on the generalized variance inflation factor (GVIF) (\citealp{fox1992generalized}). Each coefficient standard error is inversely weighted by GVIF and then summed to derive an adjusted null bound. GVIF-adjusted ProSGPV has improved inference and prediction performance than original ProSGPV when signals are dense, or correlation is high in the design matrix. The improvement is illustrated in Supplementary Figure 6 and this bound is an option in the ProSGPV package. 

In Figure \ref{Fig3}, parameter estimation bias, as measured by mean absolute error (MAE), is compared under same scenarios in Figure \ref{Fig2}. ProSGPV generally has the lowest MAE in all cases. BeSS has low bias in low-dimensional case but has high variance in parameter estimation in high-dimensional Logistic and Cox regressions. ISIS and lasso tend to have decent parameter estimation only when signals are sparse. 

In Figure \ref{Fig4}, prediction accuracy, as measured by average area under the curve (AUC) in Logistic regression and root mean square error (RMSE) in Poisson regression, is compared under combinations of $(n,p,s)$ using an independent test set. In Logistic regression, when $n>p$, most algorithms reach a very high AUC as $n/p$ passes 10; when $p>n$, all algorithms achieve similar AUC even when $p$ grows. In Poisson regression, when $n>p$ and signals are sparse, all algorithms have very high prediction accuracy. When $n>p$ and signals are dense, ISIS has higher prediction error than the other algorithms. When $p>n$, ISIS and lasso has worse prediction than ProSGPV and BeSS.

Running time of all algorithms is compared in Supplementary Figure 7. When $n>p$, lasso takes the longest time to run, likely due to the cross-validation step. ISIS usually takes the second longest time to run. BeSS can take longer to run in Cox regression when $n$ is large. ProSGPV is among the algorithms that have the shortest computation time. The spike in time when $n$ is small could be due to the convergence issue given a small sample size. When $p>n$, ISIS and lasso are more time-consuming than BeSS and ProSGPV.

\section{Real world data}
\label{sec5}

We applied ProSGPV in a real-world data example to evaluate the sparsity of its solution and prediction accuracy. Lower back pain can be caused by a variety of problems with any parts of the complex, interconnected network of spinal muscles, nerves, bones, discs or tendons in the lumbar spine. Dr. Henrique da Mota collected 12 biomechanical attributes from 310 patients, of whom 100 are normal and 210 are abnormal (Disk Hernia or Spondylolisthesis). The goal is to differentiate the normal patients from the abnormal using those 12 variables. The biomechanical attributes include pelvic incidence, pelvic tilt, lumbar lordosis angle, sacral slope, pelvic radius, degree of spondylolisthesis, pelvic slope, direct tilt, thoracic slope cervical tilt, sacrum angle, and scoliosis slope. This spine data were acquired from University of California Irvine Machine Learning Repository (\citealp{Dua:2019}) on March 4th, 2021 and is available in the ProSGPV package.

The clustering and correlation pattern of the data are shown in Supplementary Figure 8. Most features are weakly associated with the outcome, and some of them are heavily correlated with each other. To understand the sparsity of solutions from various algorithms, we repeatedly split the data set into a training (70\%) and test (30\%) set. In each of 1000 repetitions, we ran ProSGPV, lasso, BeSS, and ISIS on the training set, recorded the model size, and calculated the prediction AUC in the test set. Training model sizes and most frequent models are summarized in Supplementary Figure 9. Prediction AUC are summarized in Supplementary Figure 10. Lasso selected the greatest number of variables while ProSGPV selected a sparser model on average. Pelvic radius and degree of spondylolisthesis were selected by three out of four algorithms. All algorithms yielded similar prediction performance on the test set. In summary, ProSGPV produced a sparse model with high predictive ability.

\section{Discussion}
\label{sec6}

\subsection{When ProSGPV excels}
\label{sec6:1}

As shown from simulation studies in Section \ref{sec4}, ProSGPV can recover the true support with high probability in classic $n>p$ and high dimensional $p>n$ settings. When $p>n$, signals are generally assumed to be sparse; otherwise, to our knowledge, no known algorithms would successfully recover the true support (\citealp{fan2010sure,wang2010bridge,jiang2016variable,wang2016variable,avella2018robust,salehi2019impact,ma2020global}). ProSGPV often achieves the fastest rate of being support recovery consistent. The intuition behind its nice properties is as follows. Lasso evaluated at $\lambda_\text{gic}$ reduces the feature space to a candidate set that is very likely to contain the true support. This behavior is particularly important when $p>n$. The second-stage thresholding examines whether each variable is clinically meaningful enough to stay in the model. Some statistically significant variables may be excluded from the final model, as their effects overlap with the null zone. That explains why ProSGPV has a high power and an extremely low Type I Error rate. The successful support recovery relies on the assumption that true signals have larger effect sizes than noise variables in the data set. If that is violated, ProSGPV may miss one or two signal variables whose effects are below a clinically significant level, as measured by the null bound ($\overline{SE}$). But to be fair, no current method preforms well when that overlap exists.

ProSGPV recovers the true support with high probability and its final model does not include shrunken estimates. Therefore, the parameter estimation performance of ProSGPV tends to be excellent and superior in comparison to standard methods. We conjecture that ProSGPV enjoys oracle properties of support recovery and parameter estimation. Here, oracle properties refers to that its estimates are as good as maximum likelihood estimates when the true support is known (\citealp{fan2001variable}). Our simulations, combined with the real-world data example, provide strong evidence that ProSGPV, even though it is not optimized for prediction, yields comparable prediction performance to other standard procedures.

\subsection{When ProSGPV fails}

When signals are dense, such as in a classic logistic regression and $n$ is small, ProSGPV may select fewer signal variables than the truth, resulting in a sparse model. This could be remedied by adjusting the null bound to include more variables with smaller effect sizes, as we have shown with the GVIF adjustment. However, changing the null bound in all scenarios would needlessly affect ProSGPV’s general performance. On the flip side, ProSGPV does not select any noise variables, which implies that variables selected are very likely to be true signals (low false discovery rate).

In simulation studies not shown here, ProSGPV tended to have poor support recovery performance when true effect sizes are below a certain noise level, as suggested by Section \ref{sec6:1}. It is a common assumption that the effect size of true signals are large enough to be detectable for a successful support recovery (in general, above the noise level) (\citealp{wang2010bridge,jiang2016variable,salehi2019impact}). In addition, we observed in Logistic regression that when the true effect size is too large all algorithms had difficulty identifying the true support. That is likely due to issues with complete/quasi-complete separation (\citealp{albert1984existence,rahman2017performance}). Logistic regression maps the linear predictor of explanatory variables to a continuous probability but the outcome is binary (discrete). When $n$ is small, different combinations of variables may achieve very similar probabilities for an event. Fortunately, the prediction performance does not suffer in the high dimensional dense signal setting. 

\subsection{Remedies when data are highly correlated or signals are dense}

When the design matrix is orthogonal and variables are centered and standardized, all coefficient standard errors (SE) are the same. When data columns are correlated, SEs are inflated, resulting in larger null bound in the ProSGPV. This results in screening out more small true effects and leads to worse support recovery performance, as is shown in the logistic example in Figure \ref{Fig2}. However, this can be addressed by replacing the constant null bound with a generalized variance inflation factor (GVIF) (\citealp{fox1992generalized}) adjusted null bound. This action deflates the SEs for variables that have high correlation with other variables, reduces the size of the null bound, and consequently includes slightly more variables in the ProSGPV selection set. The GVIF-adjusted ProSGPV exhibits improvement over the original ProSGPV when data are correlated and signals are dense (see Supplementary Figure 6).

\subsection{Closing comments}

There is a rich literature in variable selection methods in Logistic regression, either in classic $n>p$ or high-dimensional $p>n$ settings, variable selection in Poisson, among other GLM, and Cox regression. Algorithms proposed for Poisson and Cox models are not easily transferrable to Logistic, or other GLM. ProSGPV appears to be a unified variable selection approach that works consistently well in linear, Logistic, Poisson, and Cox models. It reduces the feature space to a smaller candidate set and applies SGPV thresholding to select variables that are clinically meaningful. Simulation studies show that ProSGPV achieves the best support recovery and parameter estimation while not giving up much prediction performance. The ProSPGV algorithm does not depend on tuning parameters that are hard to specify. When data are highly correlated or signals are known to be dense in the data, the inference and prediction performance is improved by using a GVIF-adjusted null bound in the ProSGPV algorithm. The ProSGPV algorithm is readily adaptable (e.g., using the GVIF to modify the null bound) and is fast to compute in practice. Future endeavors could be made to characterize the rate of support recovery, parameter estimation, and prediction. The novelty of the SGPV approach is that formally incorporate estimation uncertainty into variable selection tasks. This idea readily generalizes to variable selection tasks with other data types and may lead to statistical tools for other real-world problems.

\section*{Supplementary Material}

Supplementary material is available online at
\url{http://biostatistics.oxfordjournals.org}.

\section*{Acknowledgments}

{\it Conflict of Interest}: None declared.

\bibliographystyle{biorefs}
\bibliography{refs}

\begin{figure}[!p]
\centering\includegraphics[width=\textwidth]{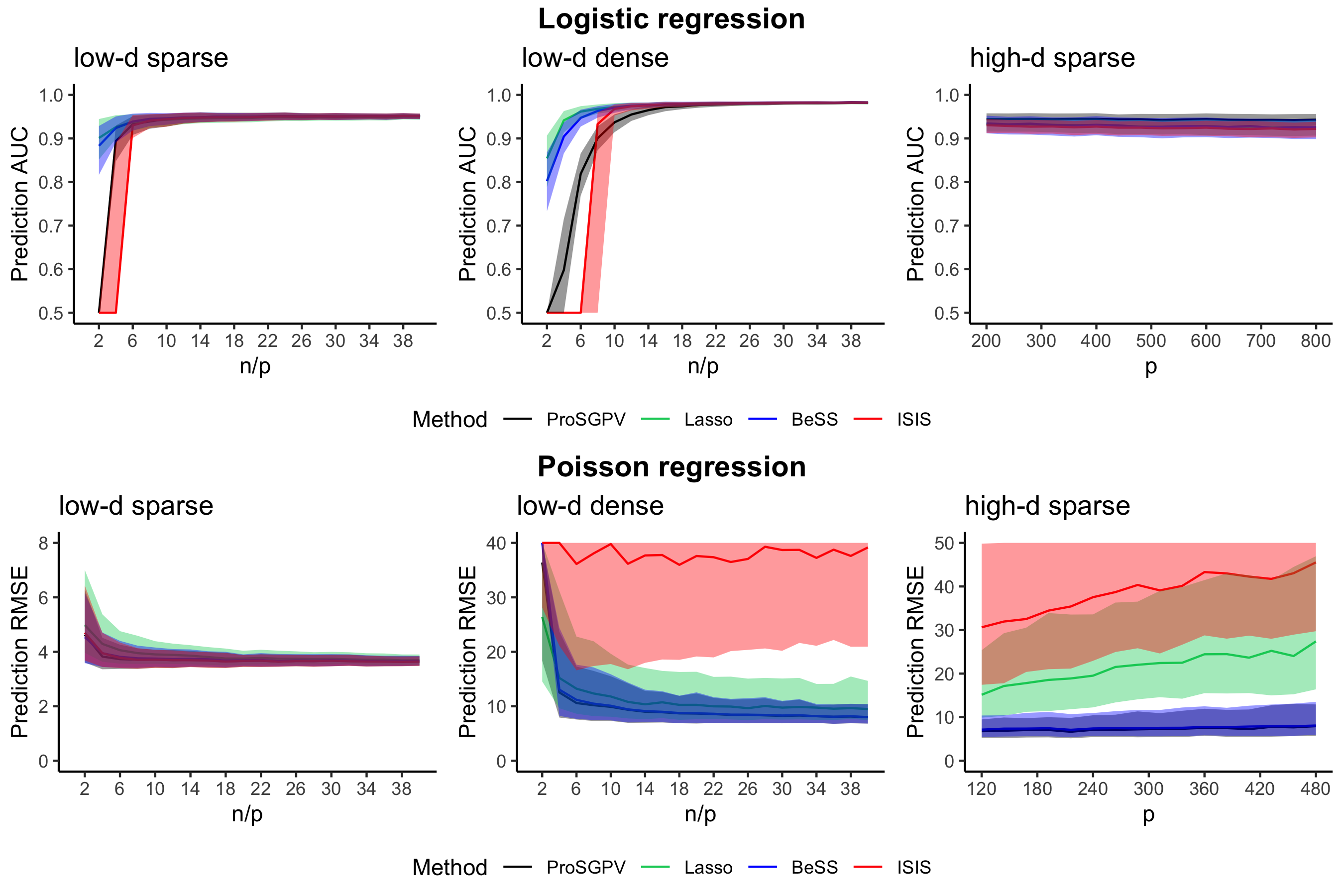}
\caption{ Comparison of prediction performance in a separate test set: median area under the curve surrounded by first and third quartiles in logistic regression and median prediction root mean square errors (RMSE) surrounded by first and third quartiles in Poisson regression. RMSE are bounded for aesthetic reasons. }
\label{Fig4}
\end{figure}

\begin{table}[!p]
\tblcaption{Summary of parameters in simulation studies. Low-s stands for low dimensional sparse; low-d stands for low dimensional dense; and high-s stands for high dimensional sparse. 
\label{Table1}}
{\tabcolsep=4.25pt
\begin{tabular}{llllllllll}
\hline
  & \multicolumn{3}{l}{Logistic regression} & \multicolumn{3}{l}{Poisson regression} & \multicolumn{3}{l}{Cox regression} \\
  & Low-s       & Low-d       & High-s      & Low-s       & Low-d      & High-s      & Low-s      & Low-d     & High-s    \\ \hline
n & 40:800      & 40:800      & 200         & 40:800      & 40:800     & 120         & 40:800     & 40:800    & 80        \\
p & 20          & 20          & 200:800     & 20          & 20         & 120:480     & 20         & 20        & 80:320    \\
s & 4           & 14          & 4           & 4           & 14         & 4           & 4          & 14        & 4         \\
$\beta_l$ & 0.5         & 0.5         & 0.5         & 0.1         & 0.1        & 0.1         & 0.2        & 0.2       & 0.2       \\
$\beta_u$ & 1.5         & 1.5         & 1.5         & 0.4         & 0.4        & 0.4         & 0.8        & 0.8       & 0.8       \\
t & 0           & 0           & 0           & 2           & 2          & 2           & 0          & 0         & 0         \\ \hline
\end{tabular}
}
\end{table}

\end{document}